\documentclass[a4paper,oneside]{saip}
\usepackage{helvet}          
\usepackage{times}           
\usepackage{courierten}   
\usepackage[T1]{fontenc} 
\usepackage{amsmath,amssymb,amsfonts,cases}
\usepackage{graphicx,caption, subcaption,booktabs}  
\usepackage{tikz}
\usetikzlibrary{positioning}
\usetikzlibrary{arrows, calc}
\usepackage{url}
\usepackage{geometry} 
\usepackage[normalem]{ulem}

\begin{document}

\title{Seed-Robust PINN Determination of $s$-Wave Bound States and Jost-Function-Based vertex constants in $_{\Lambda}^{208}$Pb}

\author{J T Tshipi$^{1}$ and A S Cornell$^{2}$}
\affil{$^1$Department of Physical and Earth Sciences, Sol Plaatje University, Kimberley, South Africa}
\affil{$^2$Department of Physics, University of Johannesburg, PO Box 524, Auckland Park 2006, South Africa}

\email{\url{tshegofatso.tshipi@spu.ac.za , acornell@uj.ac.za}}

\begin{abstract}
We investigate physics-informed neural networks (PINNs) for computing the $s$-wave bound states of the hypernucleus $_{\Lambda}^{208}$Pb, modeled as a two-body system composed of a $\Lambda$ hyperon and a $^{207}$Pb core. The interaction is described by a Woods--Saxon potential without spin--orbit coupling. In the PINN formulation, the radial bound-state wave function is represented by an artificial neural network, and the eigenenergies are obtained from the Rayleigh--Ritz variational quotient. Because PINN eigenvalue calculations can depend on the residual-loss formulation and random-seed initialization, two residual losses are compared across four independent random seeds. Their performance is assessed using eigenvalue accuracy, coefficient of variation, signal-to-noise ratio, bias--variance decomposition, and Hermitian spectral-ordering consistency. The normalized residual loss gives the most stable and physically consistent bound-state spectrum for the four seeds considered. With this loss, the computed bound-state energies and root-mean-square radii are in very good agreement with the corresponding theoretical values. The resulting wave functions are used to construct the Jost functions through the Wronskian with incoming and outgoing Riccati--Hankel functions. From these Jost functions, the residue of the partial-wave $S$-matrix at the bound-state pole, the Asymptotic Normalization and the Nuclear Vertex Constants are extracted. These quantities show reasonable agreement with theoretical values, although with larger standard deviations than those obtained for the eigenenergies and radii. The results indicate that the Rayleigh--Ritz formulation combined with a seed-robust normalized residual loss provides a stable PINN framework for bound-state and Jost-function-based calculations in hypernuclear two-body systems.
\end{abstract}

\section{Introduction}
The study of hypernuclei plays an important role in shaping our understanding of the internal structure of neutron stars, where hyperons are hypothesized to be produced under the high-pressure conditions found in the inner core~\cite{EoS_hypernucleus}. Experimentally, hypernuclei can be produced by converting a neutron into a $\Lambda$ particle through reactions such as $(\pi^+,K^+)$ and $(K^-,\pi^-)$~\cite{Hashimoto:2006aw,ZenoniGianotti2002}, or more recently through the $(e,e'K)$ reaction, where a proton is converted into a $\Lambda$ particle~\cite{208pb_papervi}.

The $\Lambda$ hyperon is the lightest hyperon and carries strangeness $S=-1$~\cite{ZenoniGianotti2002}. Owing to this strangeness degree of freedom, the $\Lambda$ hyperon can be embedded deeply inside a nucleus and can therefore act as a sensitive probe of the nuclear interior. Since the $\Lambda$ hyperon is not Pauli blocked by nucleons, it can penetrate into the nuclear interior and form deeply bound hypernuclear states~\cite{Hashimoto:2006aw}. Once the binding energies of the $\Lambda$ hyperon are known, they provide essential empirical constraints on hyperon--nucleon interactions. These interactions are important inputs in hyperonic equation-of-state (EoS) calculations and are therefore central to understanding the softening of the EoS associated with the appearance of hyperons in neutron-star matter~\cite{Vidana2018,Lonardoni2015,Zachariou2024}.

The $s$-wave binding energies of the $\Lambda$ single-particle states in $_{\Lambda}^{208}$Pb were studied in Ref.~\cite{Tshipi2025SAIP} using the unsupervised Physics-Informed Neural Network (uPINN) approach~\cite{Jin2020UnsupervisedNN}. In that earlier work, only two of the first three bound states were successfully reproduced, while the correct second excited state was not obtained. As discussed in Ref.~\cite{correction_to_usnn}, this difficulty arises because the uPINN formulation imposes an unphysical restriction on the determination of the eigenpairs during training. More specifically, the loss functions associated with the eigenfunction and eigenvalue vanish only when the eigenpairs diverge, thereby preventing convergence to the correct physical
solutions~\cite{correction_to_usnn}.

In this work, we solve the bound-state eigenvalue problem by incorporating physical constraints associated with the Hermitian nature of the bound-state Hamiltonian. For the real single-channel radial problem considered here, the bound-state energies are real, the corresponding wave functions may be chosen to be real, and the bound states are expected to be non-degenerate. Consequently, the computed eigenvalues should preserve the expected spectral ordering with increasing quantum number (Hermitian
spectral-ordering consistency)~\cite{evp_book}. In addition, the bound-state wave functions must be normalizable and mutually orthogonal. During training, the energies are approximated using the Rayleigh-Ritz variational quotient~\cite{LAGARIS19971}. The PINN is trained using residual-loss formulations designed to enforce the Schr\"odinger equation and the associated physical constraints. In particular, we compare the standard residual loss with a normalized residual loss inspired by normalized residual formulations used in applied mathematics~\cite{Rayleigh_paper}.

In addition to the $s$-wave eigenpairs, we construct the Jost functions from the Artificial Neural Network (ANN) based bound-state solutions. The Jost function plays an important role in quantum few-body problems because it provides direct access to the partial-wave $S$-matrix and its pole structure. Once the Jost function is determined, one can compute the residue of the $S$-matrix at the bound-state poles. These residues are related to the nuclear vertex constants (NVCs), which characterize the strength of the bound-state pole, and to the asymptotic normalization constants (ANCs), which determine the asymptotic behaviour of the bound-state wave function~\cite{book}. These quantities are also relevant for the description of hypernuclear peripheral radiative capture~\cite{Mmusi}. Therefore, the present study determines the $s$-wave bound states generated by a Woods-Saxon potential without spin-orbit coupling~\cite{Guleria:2011svi}, and uses the resulting bound-state solutions to calculate the corresponding residue of the $S$-matrix, ANC, and NVC within the PINN framework.

Since the Jost functions are constructed from the bound-state wave functions and their derivatives, seed-induced fluctuations in the PINN solution can propagate into the extracted residue of the $S$-matrix, ANC, and NVC. For this reason, random seeds are treated as an experimental variable in this work. The main contribution of the present study is a seed-robust residual-loss evaluation framework for PINN-based quantum bound-state calculations.

Instead of selecting a residual loss from a single training run, two residual formulations are compared across four independent random seeds. Their performance is assessed using eigenvalue accuracy, coefficient of variation, signal-to-noise ratio, bias--variance decomposition, and Hermitian spectral-ordering consistency.

The details of these seed-robustness indicators are given in Section~\ref{sec:seed_robustness_metrics}. The residual loss that produces the most stable and physically consistent bound-state spectrum is then selected for the computation of the Jost function, the residue of the $S$-matrix, the asymptotic normalization constant, and the nuclear vertex constant.

The present paper is organized as follows. In Section~\ref{sec:theory}, we briefly discuss the theory of bound states, define the Jost function, and show how the ANC, NVC, and residue of the $S$-matrix are calculated. In Section~\ref{sec:PINN}, we provide a brief summary of the PINN formulation, show how it is combined with the Rayleigh-Ritz quotient to calculate the eigenpairs, and define the loss functions used for the bound-state problem. In Section~\ref{sec:seed_robustness_metrics}, we introduce the seed-robustness metrics used to compare the residual-loss formulations. In Section~\ref{sec:result_disc}, we present and discuss the calculated single-particle energies of the $\Lambda$ states, together with the ANC, NVC, and residue of the $S$-matrix, and compare them with the theoretical results reported in Refs.~\cite{Mmusi,blokhintsev2007vertex}. Finally, in Section~\ref{sec:concl}, we summarize the main findings and present the conclusions.

\section{Theory of bound states}\label{sec:theory}
In this study, we adopt the single-particle model in which the hypernucleus $_{\Lambda}^{208}$Pb is treated as a two-body system which is made up of the $\Lambda$ particle and the $^{207}$Pb core~\cite{Mmusi,blokhintsev2007vertex}. Let $r$ denote the relative distance between the $\Lambda$ particle and the core nucleus, in which case the radial Schr\"odinger equation describing the relative motion is given by
\begin{equation}
 \label{singleChannelSE}
 \left[ -\frac{\hbar^2}{2\mu}\frac{d^2}{dr^2}+\frac{\ell\left(\ell+1 \right)}{r^2}+V(r)
 \right]u_{\ell}(E,r) = Eu_{\ell}(E,r) \; ,
\end{equation}
where
\begin{eqnarray}
    \mu = \frac{m_{Pb}m_{\Lambda}}{m_{Pb}+m_{\Lambda}}
\end{eqnarray}
is the reduced mass of the $\Lambda$ and the core nucleus. $m_{Pb}$ is the mass of $^{207}$Pb, $m_{\Lambda}$ is the $\Lambda$ particle and $E$ is the hyperon binding energy. It is important to note that $u_\ell(r)$ is the reduced radial wave function i.e.
\begin{equation*}
    u_\ell(E,r)=rR_\ell(E,r) \; .
\end{equation*}

Since we are only interested in assessing the efficiency of PINNs in solving eigenvalue problems (EVP), we restrict our attention to the $s$-wave bound-states, i.e. $\ell = 0$. The final Schr\"odinger equation~(\ref{singleChannelSE}) reduces to
\begin{equation}
 \label{singleChannelSE_SWave}
 \left[
 -\frac{\hbar^2}{2\mu}\frac{d^2}{dr^2}
 +V(r)
 \right]u_{0}(E,r)
 = Eu_{0}(E,r) \; ,
\end{equation}
or more succinctly
\begin{equation}
\label{HamiltoniansingleChannelSE_SWave}
\hat{H}u_{0}(E,r) = Eu_{0}(E,r) \; ,
\end{equation}
where $\hat{H}$ is the linear Hamiltonian operator.

For bound states, the physical solutions of Eq.~(\ref{singleChannelSE_SWave}) must satisfy
\begin{equation}
    \label{BoundaryConditions}
    u_0(E,0) = 0 \qquad \mbox{ and } \qquad u_0(E,r) \xrightarrow[r \rightarrow \infty]{} 0 \; .
\end{equation}
These conditions enforce regularity at the origin and square integrability at large distances. As a result, Eq.~(\ref{HamiltoniansingleChannelSE_SWave}) admits non-trivial bound-state solutions only for discrete values of the energy~\cite{book}.

The interaction between the $\Lambda$ hyperon and the core nucleus must be attractive and proportional to the density and size of the nucleus~\cite{Mmusi}. Moreover, the spin effects in the interaction are small and therefore can be ignored especially for heavy $\Lambda$ hyperons, like $_{\Lambda}^{208}$Pb with small values of $\ell$~\cite{Mmusi, zenoni2002physics}.

It follows that the interaction between $\Lambda$ hyperon and the core nucleus is thus described only by the Woods-Saxon potential~\cite{Mmusi, blokhintsev2007vertex,woodsax}
\begin{equation}
    \label{WoodSaxPontential}
     V(r) = - \frac{V_0}{1 + \exp\left(\frac{r-R}{d}\right)} \; ,
\end{equation}
with the parameters taken from Ref.~\cite{Mmusi},
$R = r_0(A-1)^{1/3}$ with $r_0 = 1.1$ fm and $d=0.6$ fm. The value of the strength $V_0=31.20$ MeV. Since the $\Lambda$ hyperon is electrically neutral, the Coulomb interaction is absent and is thus not included as part of the potential.

Once the wave is known, the root-mean-square radius (RMS) is calculated using
\begin{equation}
    \label{eq:rms}
    r_{\mathrm{rms}}
    \equiv
    \sqrt{\langle r^2\rangle}
    =
    \left(
    \int_0^{\infty}|u_0(E,r)|^2 r^2\,dr
    \right)^{1/2} \; .
\end{equation}

\subsection{Relationship between the wave function and Jost function}
Once the bound-state eigenpairs have been determined from Eq.~(\ref{singleChannelSE_SWave}) using the boundary conditions in Eq.~(\ref{BoundaryConditions}), the next step is to construct the Jost functions. These functions play an important role in quantum few-body problems because, once they are known, one may construct the partial-wave $S$-matrix and determine its pole structure. From the poles and residues of the $S$-matrix, the corresponding asymptotic normalization constants (ANCs) and nuclear vertex constants (NVCs) may be calculated. In this section, we focus on defining the Jost functions in terms of an ANN.

The theory of the Jost function is extensively developed in the literature~\cite{Mmusi, book,alessandrini1963jost}. Here, we highlight only the concepts relevant to our present study. Because the Woods–Saxon potential~(\ref{WoodSaxPontential}) is finite-ranged and becomes negligible at sufficiently large distances, the asymptotic behaviour of the wavefunction is governed by free incoming and outgoing spherical waves. The Jost functions are defined as the asymptotic amplitudes of these incoming, $f_\ell^{(in)}(E)$, and outgoing, $f_\ell^{(out)}(E)$, spherical waves. Mathematically, they are defined by
\begin{equation}
    \label{def_of_jost}
    f^{(in/out)}_\ell(E) = \pm \frac{i}{2k} W \left[ \chi_{\ell}^{(\pm)}(E,r),u_{\ell}(E,r) \right] \; ,
\end{equation}
where $\chi_{\ell}^{(\pm)}(E,r)$ is the Jost solution (also known as incoming $(-)$ and outgoing $(+)$ Riccati–Hankel waves),
$u_{\ell}(E,r)$ is the regular radial wave function and $W[g,f] = gf'-g'f$ is the Wronskian, which is independent of $r$ and is a non-zero constant if $f$ and $g$ are linearly independent, and zero otherwise. $f'$ and $g'$ are derivatives with respect to $r$.

The Jost solutions at asymptotic large distances are given by
\begin{equation}
    \label{JostSolution}
    \chi_{\ell}^{(\pm)}(E,r) \xrightarrow[r \rightarrow \infty]{} \left( \mp i \right)^{\ell+1}e^{\pm i kr} \; .
\end{equation}
At large distances the bound state wave function may be expressed in terms of the Jost functions,
\begin{equation}    \label{eq:wavefunction_asymptotic_with_jost}
    u_0(\mathcal{E}_d,r)  \xrightarrow[r \rightarrow \infty]{} \chi_0^{(-)}(\mathcal{E}_d,r)f^{(in)}_0(\mathcal{E}_d)
    +\chi_0^{(+)}(\mathcal{E}_d,r)f^{(out)}_0(\mathcal{E}_d)
\end{equation}

\begin{figure*}[!t]
     \centering
     \label{fig:TotalCvAndSnR}
     \begin{subfigure}[t]{\textwidth}
         \centering       \includegraphics[width=\textwidth]{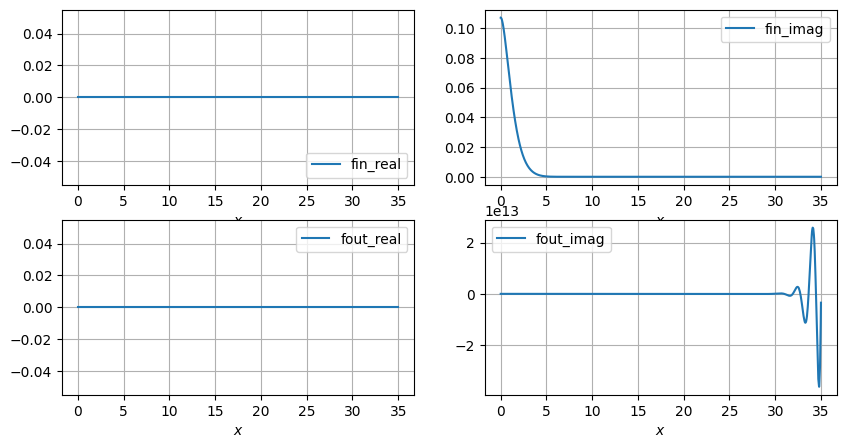}
         \caption{\it Whole domain}    \label{fig:foutgroundStateWronskianDivergin}
     \end{subfigure}
     \hfill
     \begin{subfigure}[t]{\textwidth}
     \centering
   \includegraphics[width=0.6\textwidth]{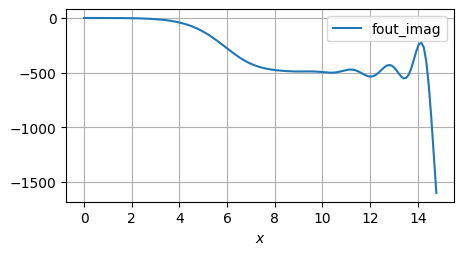}
     \caption{\it Reduced domain}
\label{fig:foutgroundStateWronskianDiverginBetter}
     \end{subfigure}
\caption{\it Incoming($f^{in}$) and outgoing($f^{out}$) Jost functions calculated from Eq.~(\ref{def_of_jost}) }
\label{fig:jostfuntionGroundState}
\end{figure*}

\noindent here, we used $k = i\kappa$ to represent the momentum of the bound state. Since the reduced $s$-wave Schrödinger equation contains no first-derivative term, Abel's identity~\cite{abelTheorem} implies that the Wronskian of $u_0(\mathcal{E}_d,r)$ and the asymptotic reference solutions $\chi_{\ell}^{(\pm)}(E,r)$ is constant only in the asymptotic radial region where the interaction potential is negligible, $V(r)\approx 0$. This is the region in which the corresponding Jost functions are extracted. At the bound state energy $E = \mathcal{E}_d$, the incoming Jost function $f^{(in)}_{\ell}(\mathcal{E}_d) \rightarrow 0$ while $f^{(out)}_{\ell}(\mathcal{E}_d)$ approaches a non-zero constant. As shown in Fig.~\ref{fig:foutgroundStateWronskianDivergin}, the computation of $f^{(\mathrm{out})}_0(\mathcal{E}_d)$ is sensitive to round-off error. For a bound state, $k=i\kappa$, so the Riccati--Hankel functions entering the Wronskian contain exponentially growing and decaying factors. At large radial distances, these factors can generate overflow or underflow, while the subtraction of nearly cancelling terms in the Wronskian can lead to loss of significant digits. These effects amplify the numerical noise observed in the computed outgoing Jost function~\cite{LanduComp}.

In Fig.~\ref{fig:foutgroundStateWronskianDiverginBetter}, we observe that reducing the computational domain drastically reduces the numerical noise associated with the calculations. The plateau region, where the Jost function is approximately constant, can be clearly seen. Because $f^{(out)}_0(\mathcal{E}_d)$ contains numerical noise, it can be estimated using a sliding-window over a narrow interval. The appropriate interval is selected using the local Signal-to-Noise Ratio (SNR) criterion, which is defined below. In each window, the local SNR is computed as a ratio between the median Wronskian magnitude and robust standard deviation estimator. Specifically, the scale of the fluctuations is estimated using the median absolute deviation (MAD), which provides a robust estimate of the standard deviation under normally distributed fluctuations and is resistant to outliers~\cite{rousseeuwCroux1993}.

At the bound state energy $\mathcal{E}_d$, the bound state wave function is related to the Jost solution through
\begin{equation}
    \label{defANC}
    u_\ell(\mathcal{E}_d,r) = i^{\ell +1}\mathcal{A}_\ell\chi^{(+)}_{\ell}(\mathcal{E}_d,r) \; ,
\end{equation}
here $\mathcal{A}_\ell$ is the Asymptotic Normalization Constant(ANC).

For a simple bound-state pole at $E=\mathcal{E}_d$, the residue of the $S$-matrix
\begin{eqnarray}
    \label{smatrix}
    S_{\ell}(E) =   f^{(out)}_\ell(E)\left[f^{(in)}_\ell(E)\right]^{-1}
\end{eqnarray}
is given by
\begin{equation}
    \label{residue_S_matrix}
    \mathrm{Res}\left[ S_{\ell}, \mathcal{E}_d \right]
    = \frac{ f^{(out)}_\ell(E)}{ \dot{f}^{(in)}_\ell(E)}
    = (-1)^{\ell+1}\frac{ik_d}{\mu} \mathcal{A}_\ell^2 \; .
\end{equation}
Here, $\dot{f}^{(in)}_\ell(E)$ represents a derivative with respect to the energy.

The Nuclear Vertex Constant (NVC) is obtained from
\begin{equation}
    \label{Nuclear_vertex_cons}
    G^2 = (-1)^{\ell} \frac{i\pi}{\mu k_d} \mathrm{Res}\left[ S_{\ell}(\mathcal{E}_d), \; \mathcal{E}_d \right] \; .
\end{equation}
Eqs.~(\ref{def_of_jost})-(\ref{Nuclear_vertex_cons}) show that once the bound-state wave function is known and the Jost functions have been constructed, the residue of the $S$-matrix, the ANC, and the NVC may be calculated within a common framework.

\section{Physics Informed Neural Network}\label{sec:PINN}
The Physics-informed Neural Network (PINN) approach~\cite{Raissi} is used to solve the $s$-wave bound-state eigenvalue problem defined in
Eq.~(\ref{HamiltoniansingleChannelSE_SWave}). Prior to imposing the bound-state spectral condition, the radial Schrödinger equation may be
regarded as admitting an energy-parametrised solution, denoted by $u_0(E,r)$. For bound states satisfying the required boundary conditions
and normalisability condition, physical solutions occur only at discrete eigenenergies $E_n$, so that the corresponding radial eigenfunction is
$u_{0,n}(r)\equiv u_0(E_n,r)$. In the present PINN spectral formulation, the neural network is therefore used to approximate the
$n^{\text{th}}$ bound-state eigenfunction rather than the full energy-dependent family of solutions. Thus,
\begin{equation}
u_{0,n}(r) \equiv u_0(E_n,r) \approx \phi_n(r;\theta) \; ,
\end{equation}
where $\phi_n(r;\theta)$ denotes the ANN approximation to the $n^{\text{th}}$ bound-state wave function in the computational domain.
Rather than treating the energy as an independent trainable parameter, the corresponding approximate eigenvalue is determined self-consistently
from the trial wave function through the Rayleigh-Ritz variational quotient~\cite{LAGARIS19971,Rampho2009_thesis},
\begin{equation}
    \label{variational_E}
    \tilde{E}_n[\phi_n]
    =
    \frac{\langle \phi_n,\hat{H}\phi_n\rangle}{\langle \phi_n,\phi_n\rangle}
    =
    \frac{\displaystyle\int_{r_0}^{r_N}\phi_n^*(r;\theta)\hat{H}\phi_n(r;\theta)\,dr}
         {\displaystyle\int_{r_0}^{r_N}\left|\phi_n(r;\theta)\right|^2\,dr} \; .
\end{equation}

Since the Hamiltonian $\hat{H}$ is Hermitian on the chosen domain, the Rayleigh--Ritz quotient is real for admissible nonzero trial functions
satisfying the required boundary conditions. For normalized trial functions, it is also bounded from below by the ground-state energy of
$\hat{H}$. This provides a physically constrained estimate of the eigenvalue and helps stabilize the PINN eigenvalue calculation during
training.

Within this spectral PINN formulation, the explicit energy argument is removed from the neural-network ansatz; the energy is instead computed
self-consistently as a functional of the trial wave function. The bound-state eigenvalue problem has real eigenpairs, with wave
functions that are normalizable and mutually orthogonal for distinct eigenstates. These physical properties are incorporated into the
composite loss function
\begin{equation}
    \label{total_Loss_function}
    L(\theta)
    =
    w_{\mathrm{res}}L_{\mathrm{res}}(\theta)
    +w_{\mathrm{bc}}L_{\mathrm{BC}}(\theta)
    +w_{\mathrm{orth}}L_{\mathrm{orth}}(\theta)
    +w_{\mathrm{norm}}L_{\mathrm{norm}}(\theta) \; ,
\end{equation}
which is minimized using gradient-based optimizers to obtain the optimal network parameters.

The differential-equation residual loss is defined as
\begin{eqnarray}
    \label{residue_loss}
    L_{\mathrm{res}}(\theta)
    =
    \frac{1}{N-1}\sum_{i=1}^{N-1}
    \left(
    \left[
    -\frac{\hbar^2}{2\mu}\frac{d^2}{dr^2}
    +V(r_i)
    \right]\phi_n(r_i;\theta)
    -\tilde{E}_n[\phi_n]\phi_n(r_i;\theta)
    \right)^2 \; ,
\end{eqnarray}
where the residual is evaluated at the interior collocation points $r_i$, $i=1,\dots,N-1$, and $n$ labels the bound state. Because
$\tilde{E}_n$ is itself a functional of $\phi_n$, the residual term is coupled to the Rayleigh--Ritz estimate during training. This coupling is
deliberate in the present spectral formulation, although it can increase the stiffness of the optimization problem.

The boundary loss enforces the bound-state conditions at the endpoints of the truncated computational domain,
\begin{equation}
    \label{bc_loss}
    L_{\mathrm{BC}}(\theta)
    =
    \left[\phi_n(r_0;\theta)\right]^2
    +
    \left[\phi_n(r_{\max};\theta)\right]^2 \; ,
\end{equation}
where $r_0=0$ and $r_{\max}$ is chosen sufficiently large so that the asymptotic bound-state condition is well approximated at the outer
boundary.

The normalization loss,
\begin{eqnarray}
    \label{norm_loss}
    L_{\mathrm{norm}}(\theta)
    =
    \left(
    \langle \phi_n(r;\theta),\phi_n(r;\theta)\rangle - 1
    \right)^2 \; ,
\end{eqnarray}
prevents the network from converging to the trivial zero solution.

For excited states, orthogonality is enforced through an orthogonality penalty. If the states are obtained sequentially, a natural choice is
\begin{equation}
    \label{orth_loss}
    L_{\mathrm{orth}}(\theta)
    =
    \sum_{m=0}^{n-1}
    \left(
    \langle \phi_n(r;\theta),\phi_m(r)\rangle
    \right)^2 \; ,
\end{equation}
where $\phi_m(r)$ denotes the previously computed lower-lying bound-state wave functions. This sequential strategy introduces a dependence on the
order in which the states are trained, since errors in lower-lying states can propagate to higher states through the orthogonality penalty. This
effect is one possible contributor to seed sensitivity in excited-state calculations.

The residual contribution, combined with the nonlinear neural-network parametrization, leads to a highly non-convex optimization landscape with
multiple local minima and saddle points, and may therefore be ill-conditioned~\cite{KIYANI2025118308,rathore2024challengestrainingpinnsloss}.
Moreover, the magnitude of the unnormalized residual can vary widely during training and may dominate the composite objective, making the
choice of loss weights sensitive.

To constrain the contribution of the residual loss and improve conditioning, we replace Eq.~(\ref{residue_loss}) by a normalized residual inspired by Ref.~\cite{Rayleigh_paper},
\begin{equation}
 \label{residue_loss_relative_corrected}
    L_{\mathrm{res}}(\theta)
    =
    \left(
    \frac{\|r(\phi_n)\|_2^2}
         {\|\hat{H}\phi_n\|_2^2+\|\tilde{E}_n[\phi_n]\phi_n\|_2^2}
    \right)^2 \; ,
\end{equation}
where
\begin{eqnarray}
\label{residual_equation}
    r(\phi_n) = \hat{H}\phi_n-\tilde{E}_n[\phi_n]\phi_n
\end{eqnarray}
is the differential-equation residual. This normalized form was numerically observed to improve stability across all tested seeds. To obtain the optimal ANN parameters, training is carried out in two stages: first with the Adam optimizer, followed by L-BFGS refinement. Since the primary objective is to determine eigenpairs that satisfy Eq.~(\ref{HamiltoniansingleChannelSE_SWave}), the residual component is assigned a larger weight $w_{\mathrm{res}}$ in Eq.~(\ref{total_Loss_function}). This encourages the ANN to prioritize the differential-equation constraint before refining the remaining physical constraints, thereby reducing the risk of convergence to suboptimal solutions~\cite{weight_terms}.


\section{Seed-robustness metrics for residual-loss selection}
\label{sec:seed_robustness_metrics}

The Jost function, the residue of the $S$-matrix, the asymptotic normalization constant (ANC),
and the nuclear vertex constant (NVC) all depend on a bound-state wave function that should be stable
under random seed variation. This is because the Jost functions in Eq.~(\ref{def_of_jost}) are constructed from the regular radial wave function and its derivative through the Wronskian. Consequently, seed-induced fluctuations in the PINN wave function can propagate into
$f^{(\mathrm{in})}_\ell(E)$ and $f^{(\mathrm{out})}_\ell(E)$ and, subsequently, into the extracted
residue, ANC, and NVC.

Therefore, before proceeding to the Jost-function stage, we compare the standard residual loss in Eq.~(\ref{residue_loss})
and the normalized residual loss in Eq.~(\ref{residue_loss_relative_corrected}) using seed-robustness indicators. The aim is to select the loss formulation that gives the smallest seed-to-seed variability and the smallest systematic deviation from the theoretical eigenvalues given in literature, while preserving the expected physical level ordering $E_0<E_1<E_2$ for the non-degenerate bound states. Note that random seeds are treated here as an experimental variable since neural-network training can be sensitive to initialization, sampling and the non-convex optimization landscape~\cite{Bethard2022,Neal2018,Adlam2020}. In the present PINN calculations, changing the seed affects the initial weights and the distribution of collocation points when Latin Hypercube Sampling is used.

Because eigenvalue calculations with PINNs are computationally expensive, four random seeds are used. For each bound level $n$ ($n=0,1,2$), we collect eigenvalues $E_n^{(s_1)},E_n^{(s_2)},E_n^{(s_3)},E_n^{(s_4)}$ and define the seed mean $\bar{E}_n$ and sample standard deviation $\sigma_n$ as
\begin{equation}
    \bar{E}_n =\frac{1}{N_s}\sum_{s=1}^{N_s}E_n^{(s)},
    \qquad
    \sigma_n=
    \sqrt{
    \frac{1}{N_s}
    \sum_{s=1}^{N_s}\left(E_n^{(s)}-\bar{E}_n\right)^2 } .
    \label{eq:mean_std_seed}
\end{equation}

The coefficient of variation and signal-to-noise ratio are defined as
\begin{equation}
\mathrm{CV}_n = \frac{\sigma_n}{|\bar{E}_n|}, \qquad
\mathrm{SNR}_n = \frac{|\bar{E}_n|}{\sigma_n}.
\label{eq:cv_snr_def}
\end{equation}
These are standard and well-defined.

While CV and SNR quantify relative spread, they do not distinguish systematic bias from stochastic variability. To address this, we use the bias-variance decomposition:
\begin{equation}
MSE_n = \frac{1}{N_s}\sum_{s=1}^{N_s}\bigl(E_n^{(s)} - E_{\mathrm{ref},n}\bigr)^2
= \bigl(\bar{E}_n - E_{\mathrm{ref},n}\bigr)^2
+ \frac{1}{N_s}\sum_{s=1}^{N_s}\bigl(E_n^{(s)}-\bar{E}_n\bigr)^2
\equiv \mathrm{Bias}_n^2 + \mathrm{Var}_n .
\label{eq:bias_variance_eigenvalues}
\end{equation}
The preferred residual loss is therefore the one that simultaneously gives low bias, low variance, low CV, high SNR, and preserves spectral ordering $E_0<E_1<E_2$.

\section{Results and Discussion}\label{sec:result_disc}

\begin{table*}[t!]
\centering
\caption{\it Comparison of $s$-wave bound-state eigenvalues for the Woods-Saxon potential, obtained from four independent training runs initialized with random seeds 1234, 5678, 9101, and 1112. The results were computed using the residual loss term in Eq.~(\ref{residue_loss}).}
\label{tab:eigenvalue_seed_comparison}
\begin{subtable}[t]{0.48\textwidth}
\centering
\caption{\it Eigenvalues obtained using random seeds 1234, 9101, and 1112, where the predicted bound states are not correctly ordered.}
\label{tab:wrong_table_seeds}
\begin{tabular}{ccc}
\toprule[1.5pt]
$n$ & $E$ (MeV) & Ref. \\
\midrule
$0$ & $-27.00$ & \cite{blokhintsev2007vertex} \\
    & $-27.00085$ & \cite{Mmusi} \\
    & $0.22962$ & This work \\
\midrule
$1$ & $-16.43$ & \cite{blokhintsev2007vertex} \\
    & $-16.42311$ & \cite{Mmusi} \\
    & $-27.00087$ & This work \\
\midrule
$2$ & $-3.41$ & \cite{blokhintsev2007vertex} \\
    & $-3.40020$ & \cite{Mmusi} \\
    & $-16.42331$ & This work \\
\bottomrule[1.5pt]
\end{tabular}
\end{subtable}
\hfill
\begin{subtable}[t]{0.48\textwidth}
\centering
\caption{\it Eigenvalues obtained using random seed 5678, showing the correct ordering of the bound states.}
\label{tab:right_table_seed}
\begin{tabular}{ccc}
\toprule[1.5pt]
$n$ & $E$ (MeV) & Ref. \\
\midrule
$0$ & $-27.00$ & \cite{blokhintsev2007vertex} \\
    & $-27.00085$ & \cite{Mmusi} \\
    & $-27.00094$ & This work \\
\midrule
$1$ & $-16.43$ & \cite{blokhintsev2007vertex} \\
    & $-16.42311$ & \cite{Mmusi} \\
    & $-16.42320$ & This work \\
\midrule
$2$ & $-3.41$ & \cite{blokhintsev2007vertex} \\
    & $-3.40020$ & \cite{Mmusi} \\
    & $-3.40045$ & This work \\
\bottomrule[1.5pt]
\end{tabular}
\end{subtable}

\vspace{0.3cm}
\end{table*}

Throughout this study, we used an ANN with three hidden layers where each layer had 32 nodes with a $\sin(\cdot)$ activation function. This activation function has been shown to significantly accelerate the network’s convergence to a solution~\cite{Jin2020UnsupervisedNN}. Training was performed in two stages. We first used the Adam optimizer with a learning rate of $10^{-3}$, and then switched to the L-BFGS optimizer after $10^4$ iterations. A total of 5000 collocation points were generated by Latin Hypercube Sampling (LHS) over the domain $[0,35]$ fm. All integrals were evaluated using Gauss--Legendre quadrature with 512 points. The loss weights were chosen as $w_{\mathrm{res}}=150$, while the remaining weights were set equal to unity. All computations were carried out in \texttt{PyTorch}.

Table~\ref{tab:eigenvalue_seed_comparison} shows the $s$-wave bound-state eigenvalues obtained using the residual loss term in Eq.~(\ref{residue_loss}), where both the ANN and LHS were initialized using four different random seeds. The results in Table~\ref{tab:wrong_table_seeds}, obtained using seeds 1234, 9101, and 1112, show an incorrect ordering of the predicted bound states. In particular, the state labelled as $n=0$ gives a positive, spurious energy, while the energies corresponding to the ground and first excited states appear to be shifted to $n=1$ and $n=2$, respectively. The expected second excited state at $E=-3.4$ MeV is not recovered. In contrast, the result obtained using seed 5678, shown in Table~\ref{tab:right_table_seed}, reproduces the expected ordering and agrees well with theoretical values. These observations indicate that the residual loss in Eq.~(\ref{residue_loss}) can yield accurate eigenvalues, but its performance is strongly dependent on the choice of initialization seed.

\begin{figure*}[h!]
     \centering
     \begin{subfigure}[t]{0.8\textwidth}
         \centering        \includegraphics[width=\textwidth]{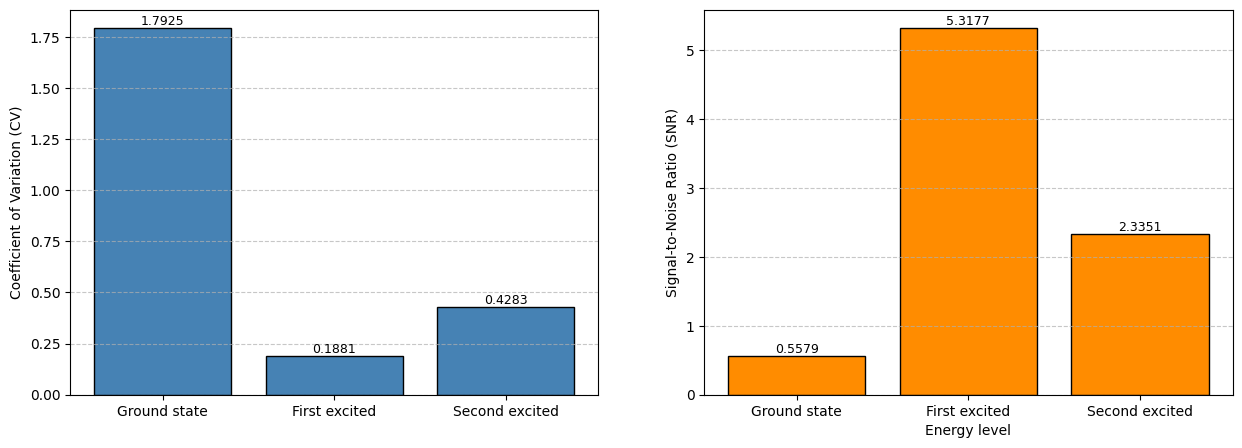}
         \caption{\it Residual loss}    \label{fig:CVAndSnRNormalResidue}
     \end{subfigure}
     \hfill
     \begin{subfigure}[t]{0.7\textwidth}
     \centering
     \includegraphics[width=\textwidth]{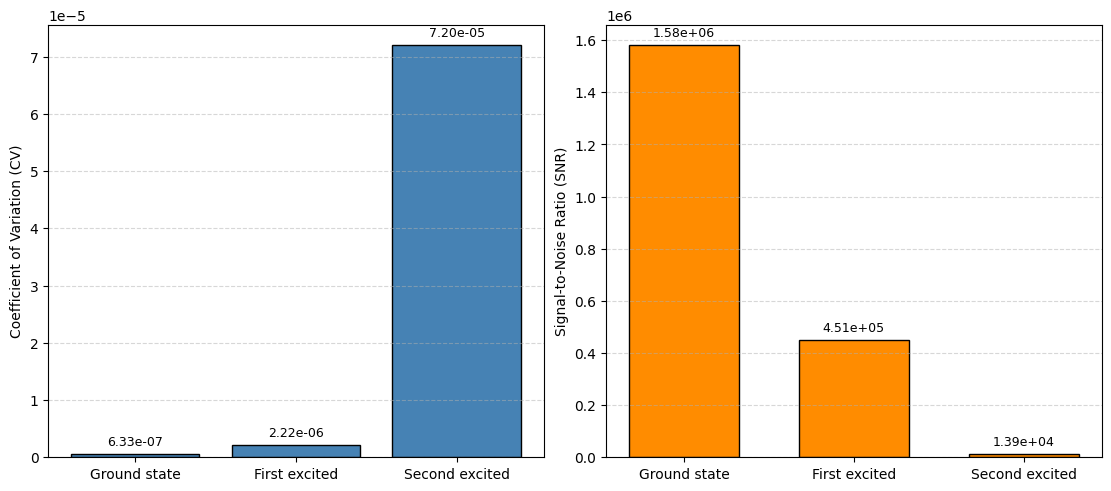}
     \caption{\it Normalized residual loss}
   \label{fig:CVAndSnRNormalizedResidue}
     \end{subfigure}
     \caption{\it CV and the corresponding SNR using Eq.~(\ref{residue_loss}) and normalized residual Eq.~(\ref{residue_loss_relative_corrected}) losses.  }   
     \label{fig:TotalCvAndSnR}
\end{figure*}

To address this sensitivity, the normalized residual loss defined in Eq.~(\ref{residue_loss_relative_corrected}) is employed. The corresponding results, shown in Table~\ref{tab:meanEigenvaluesandstd}, demonstrate that the predicted eigenvalues become stable across all four initialization seeds. This improvement is quantified using the coefficient of variation (CV) and the signal-to-noise ratio (SNR), as shown in Fig.~\ref{fig:TotalCvAndSnR}. From Fig.~\ref{fig:CVAndSnRNormalResidue}, the unnormalized residual loss yields CV values greater than zero for all states, indicating sensitivity to seed variation. The ground-state energy exhibits the largest variation, while the first excited state is less affected, consistent with its larger SNR and smaller relative error. This behaviour correlates with the presence of a spurious positive-energy solution.

\begin{figure*}[h!]
     \centering
     \begin{subfigure}[t]{0.8\textwidth}
         \centering
    \includegraphics[width=\textwidth]{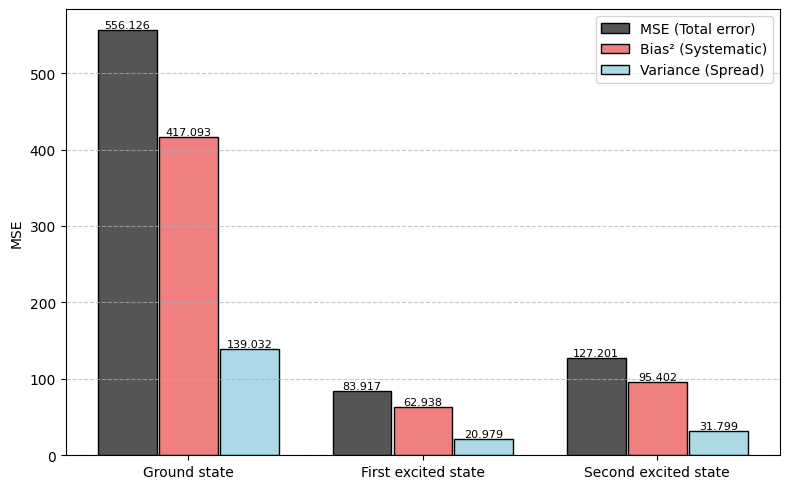}
         \caption{\it Residual loss}    \label{fig:BiaseAndVarianceNormalResidue}
     \end{subfigure}
     \hfill
     \begin{subfigure}[t]{0.8\textwidth}
     \centering
     \includegraphics[width=\textwidth]{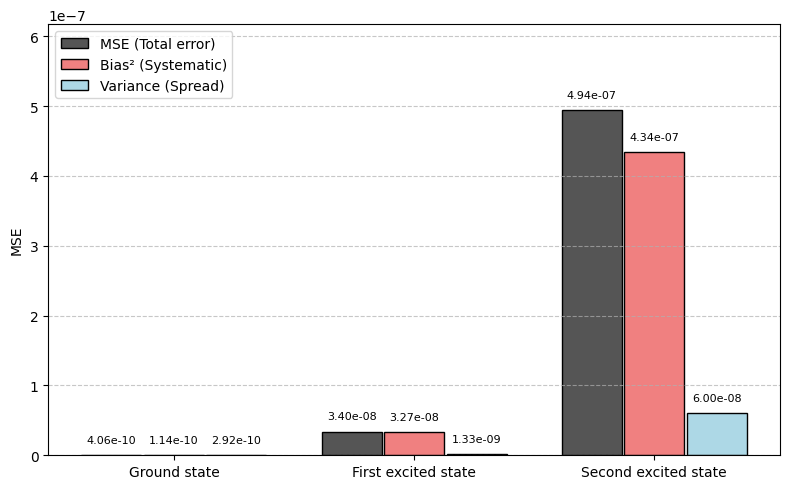}
     \caption{\it Normalized residual loss}
\label{fig:BiaseAndVarianceNormalizedResidue}
     \end{subfigure}
     \caption{\it Bias-Variance plots for the residual Eq.~(\ref{residue_loss}) and normalized residual Eq.~(\ref{residue_loss_relative_corrected}) losses.  }   
     \label{fig:BiasVarianceFigs}
\end{figure*}

In contrast, Fig.~\ref{fig:CVAndSnRNormalizedResidue} shows that the normalized residual loss produces CV values close to zero and significantly larger SNR values across all states. These results indicate that the normalized residual loss substantially reduces seed dependence and stabilizes the predicted eigenvalues. To further understand this behaviour, the bias–variance decomposition is examined. In Fig.~\ref{fig:BiaseAndVarianceNormalResidue}, the bias squared is approximately three times larger than the variance, indicating a high-bias, low-variance regime. This shows that the unnormalized residual loss systematically converges to nonphysical solutions, producing eigenvalues that are consistent but incorrect. In contrast, Fig.~\ref{fig:BiaseAndVarianceNormalizedResidue} shows that both the bias squared and variance are reduced to values close to zero when the normalized residual loss is used. This indicates that the corresponding eigenvalues are both precise and accurate, and that the solution is robust with respect to initialization. It is further observed that the normalized residual loss improves the learning of the ground-state eigenvalue, as both the bias squared and variance attain their smallest values for this state. For the excited states, the bias squared term dominates the residual error, similar to the behaviour observed with the unnormalized loss, but at significantly reduced magnitudes. The convergence of four independent initializations to the same physically correct eigenvalues suggests that the PINN formulation with the normalized residual loss is characterized by a dominant minimum corresponding to the true Schrödinger bound-state spectrum.

Based on these observations, the normalized residual loss is adopted for all subsequent calculations. In particular, it is used for the computation of the Jost functions, as well as the residue of the $S$-matrix, the asymptotic normalization coefficient (ANC), and the nuclear vertex constant (NVC). This ensures that all derived scattering observables are based on a bound-state spectrum that is both numerically stable and physically consistent.

\begin{table}[ht!]    
\centering
 \caption{\it Mean values and standard deviations of the root-mean-square radii for the $s$-wave bound states, computed over four training runs initialized with random seeds 123, 456, 789, and 1011.} 
 \label{tab:rmsForAllseeds}
 \begin{tabular}{@{}*5c@{}}          
\toprule[1.5pt] 
\multicolumn{1}{c}{Sub shell} & \multicolumn{1}{c}{$r_{rms}$ (fm)}&\multicolumn{1}{c}{Ref}\\             
\bottomrule[1.5pt]    
1s&
 $3.58$&
 \cite{blokhintsev2007vertex}\\
 &
$ 3.5755987 \pm 2.27\times 10^{-5} $&
 This work\\   
 2s&
 $4.36$&
 \cite{blokhintsev2007vertex}\\
 &
$4.3630880\pm 6.724\times 10^{-5}$&
This work\\ 
 3s&
 $5.92$&
 \cite{blokhintsev2007vertex}\\&
$ 5.9270186 \pm 3.851\times 10^{-4}$&
This work\\  
 \bottomrule[1.5pt]    
\end{tabular}
\end{table}

In addition to the eigenvalues, the spatial properties of the wave functions are examined through the RMS radii. Table~\ref{tab:rmsForAllseeds} shows that the RMS radii, averaged over four initialization seeds, are in close agreement with the corresponding reference values, with very small standard deviations. This provides further evidence that the normalized residual loss yields stable and physically meaningful wave functions.

A more detailed examination shows that for the $1s$ state, the predicted rms radius closely matches the reference value, with deviations well within the reported standard deviation. Similar agreement is observed for the $2s$ state, where the predictions remain consistent across all seeds. For the $3s$ state, the agreement is also maintained, although with a slightly larger spread compared to the lower-lying states. Overall, the small standard deviations indicate that the spatial structure of the wave functions is robust with respect to initialization, confirming that both eigenvalues and eigenfunctions are reliably captured.

\begin{figure}[h!]
\centering
\includegraphics[width=1\textwidth]{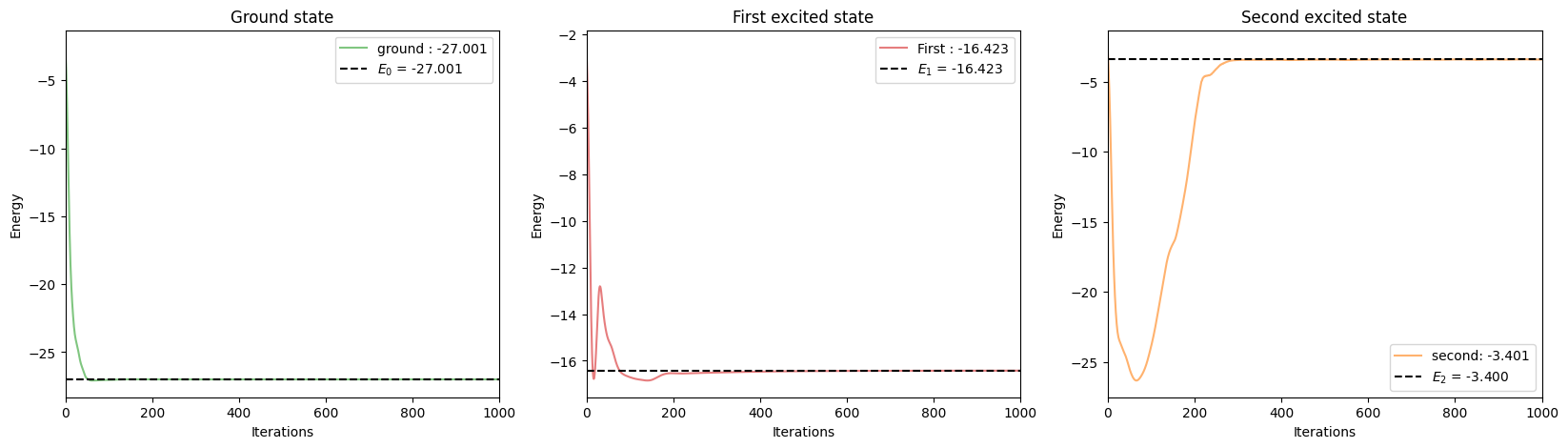}
\caption{\it Training history of the computed $s$-wave eigenvalues. The dashed horizontal lines denote the theoretical values, while the solid curves represent the ANN predictions.}
\label{fig:eigenvalueHistory}
\end{figure}

The convergence behaviour of the method is illustrated in Fig.~\ref{fig:eigenvalueHistory}. The eigenvalues obtained from the Rayleigh–Ritz quotient converge within fewer than $10^3$ iterations out of the total $10^5$ training iterations for all three $s$-wave bound states, demonstrating rapid convergence. Beyond this point, the predicted energies remain unchanged, and this regime is therefore not shown. The convergence histories for the remaining initialization seeds are not included, as they exhibit similar behaviour.

\begin{figure*}
\centering
\includegraphics[width=\textwidth]{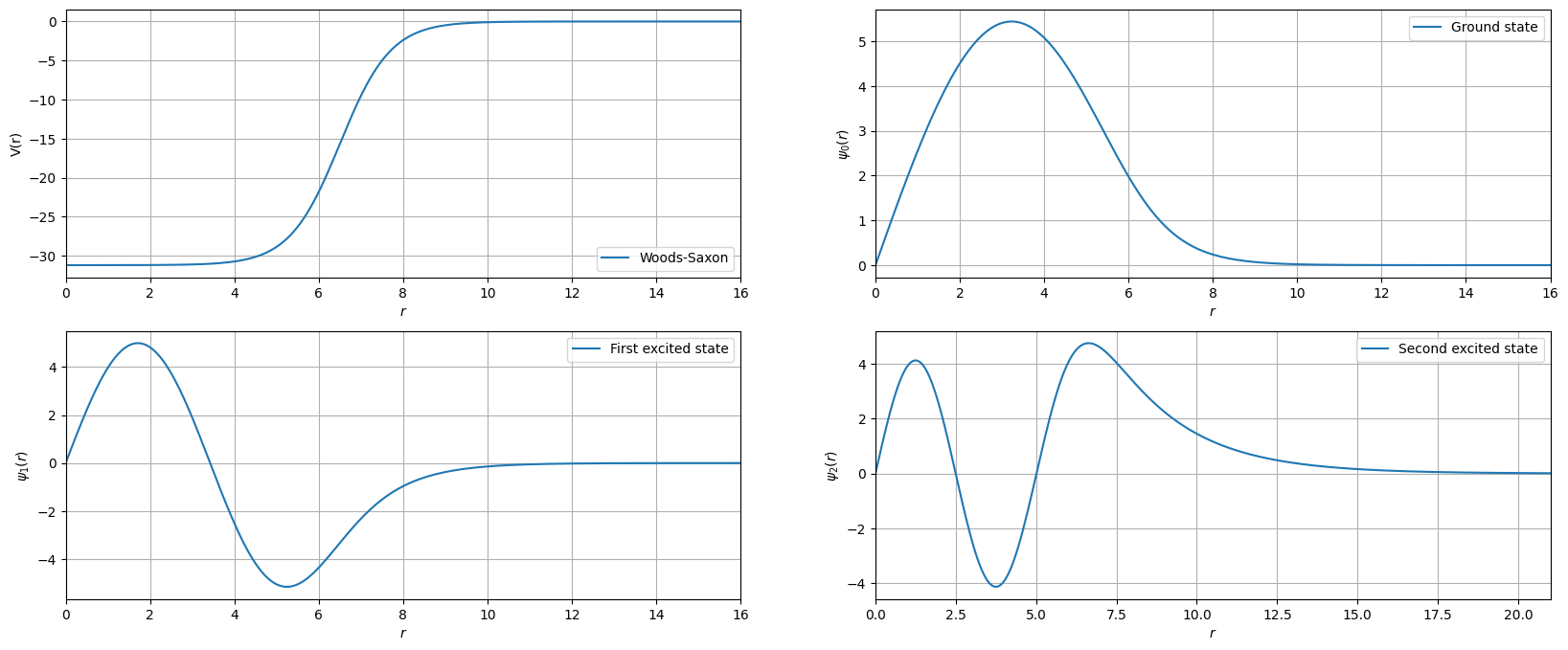}
\caption{\it Woods-Saxon potential together with the three lowest $s$-wave bound-state wave functions for $\ell=0$.}
\label{fig:WaveFunsWoods}
\end{figure*}

Finally, Fig.~\ref{fig:WaveFunsWoods} shows the Woods–Saxon potential together with the corresponding $s$-wave bound-state wave functions. The wave functions exhibit the expected nodal structure, with the number of nodes increasing with excitation level, consistent with standard quantum-mechanical behaviour. This further confirms that the proposed approach accurately reproduces both the energy spectrum and the associated wave functions.

\begin{table}[t!]
 \begin{center}
   \caption{\it Mean values and standard deviations of the $s$-wave energies, residues of the $S$-matrix, NVCs, and ANCs for the Woods--Saxon potential, obtained using the normalized residual loss in Eq.~(\ref{residue_loss_relative_corrected}) over four training runs initialized with random seeds 1234, 5678, 9101, and 1112. Results are reported as $\mu \pm \sigma$} 
   \label{tab:meanEigenvaluesandstd}
 \begin{tabular}{@{}*6c@{}}    
\toprule[1.5pt] 
\multicolumn{1}{c}{State} & \multicolumn{1}{c}{$E$ (MeV)}& \multicolumn{1}{c}{Res$\left[ s_0,\mathcal{E}\right] (\mbox{fm}^{-1})$}&\multicolumn{1}{c}{$G^2$(fm)}&\multicolumn{1}{c}{$\left| \mathcal{A}_0\right| (\mbox{fm}^{-1/2}) $}&\multicolumn{1}{c}{Ref.}\\             
\bottomrule[1.7pt]    
$1s$&
 $-27.00$&
 -&
24213.63&
493.74&
 \cite{blokhintsev2007vertex}\\
 &
$ -27.00085$&
  53546.56&
  24126.79&
  492.63&
 \cite{Mmusi} \\  
  & 
 $-27.000874 \pm 2.3\times10^{-5}$&
$ 53326.93 \pm 232.72 $&
 $24027.51\pm 104.86$ &
 $491.62\pm 1.07$&
 This work\\ \\   
\newline 
 $2s$&
 $-16.43$&
 -&
 5165.33&
 228.04&
 \cite{blokhintsev2007vertex}\\
 &
$ -16.42311$&
  8903.42&
  5143.82&
  227.47&
\cite{Mmusi}\\
 \newline
 &
 $ -16.423280 \pm 6.7\times10^{-5}$&
$8878.00 \pm 7.33$&
$5129.034\pm 4.24$&
$227.14\pm 0.09$&
 This work\\ \\
\newline 
 $3s$&
 $-3.41$&
 -&
 13.17&
 11.52&
 \cite{blokhintsev2007vertex}\\&
$ -3.40020$&
  10.965&
  13.922&
  11.834&
\cite{Mmusi}\\  
 \newline
 &
 $-3.4006395 \pm 3.851\times 10^{-4} $&
$10.960 \pm 0.005$&
$13.915\pm 0.006$&
$11.831\pm 0.002$&
 This work\\   
 \bottomrule[1.5pt]    
\end{tabular}
\end{center} 
\end{table}  

\begin{figure*}[h!]
 \centering
 \includegraphics[width=\textwidth]{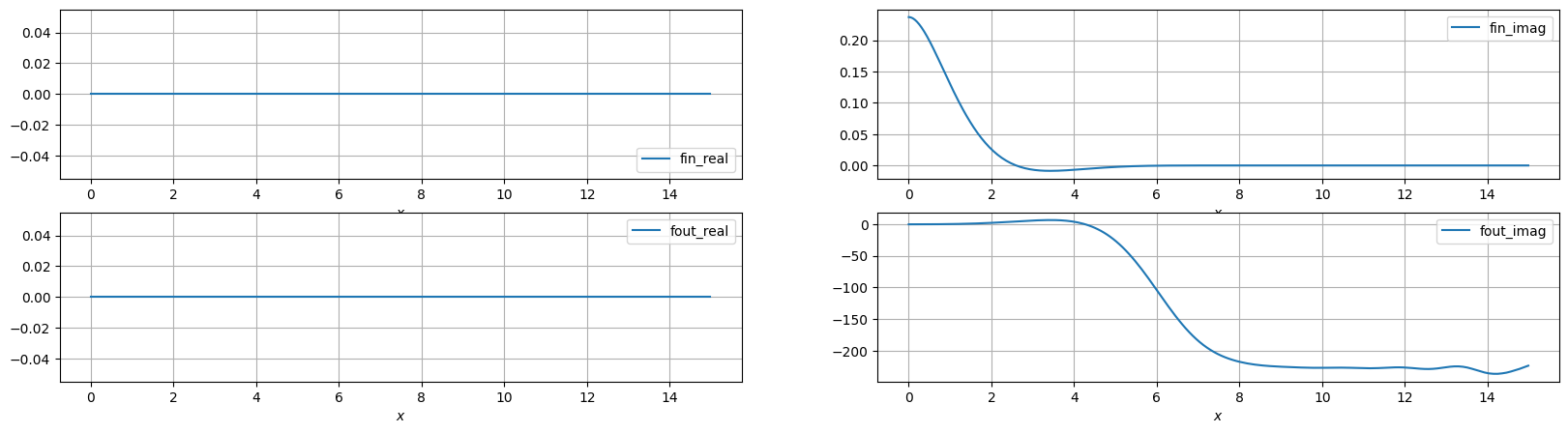}
\caption{\it Incoming (fin) and outgoing (fout) Jost functions calculated from Eq.~(\ref{def_of_jost}) for the first excited state. }   
 \label{fig:finfoutFirstExcitedState}
\end{figure*}

\begin{figure*}[h!]
 \centering
 \includegraphics[width=\textwidth]{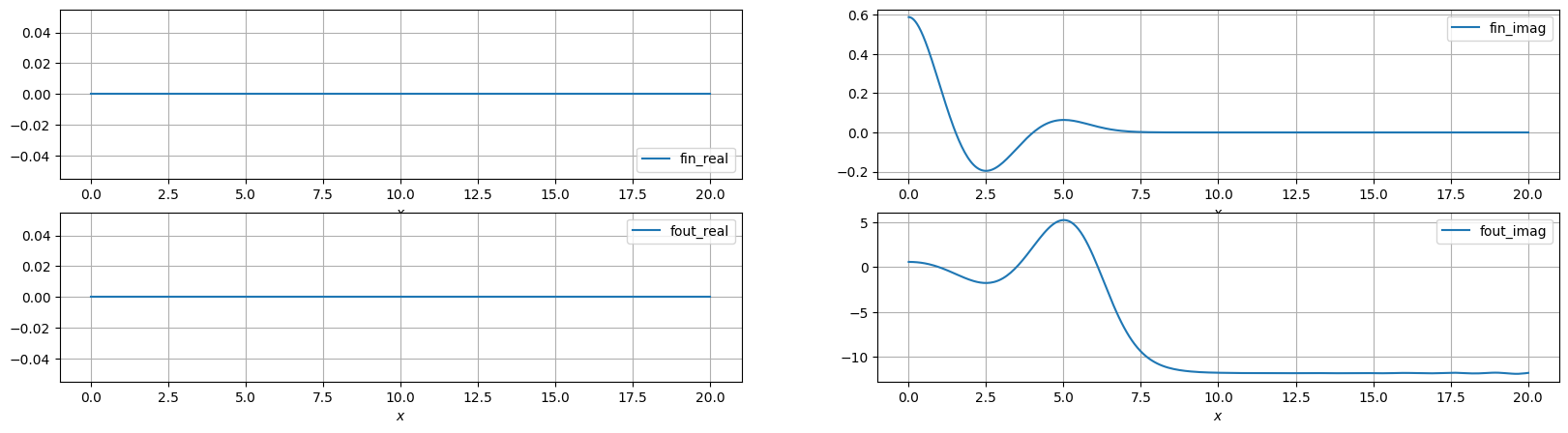}
\caption{\it Incoming (fin) and outgoing (fout) Jost functions calculated from Eq.~(\ref{def_of_jost}) for the second excited state.} 
\label{fig:finfoutSecondExcitedState}
\end{figure*}

\begin{figure*}[h!]
\centering
\includegraphics[width=0.85\textwidth]{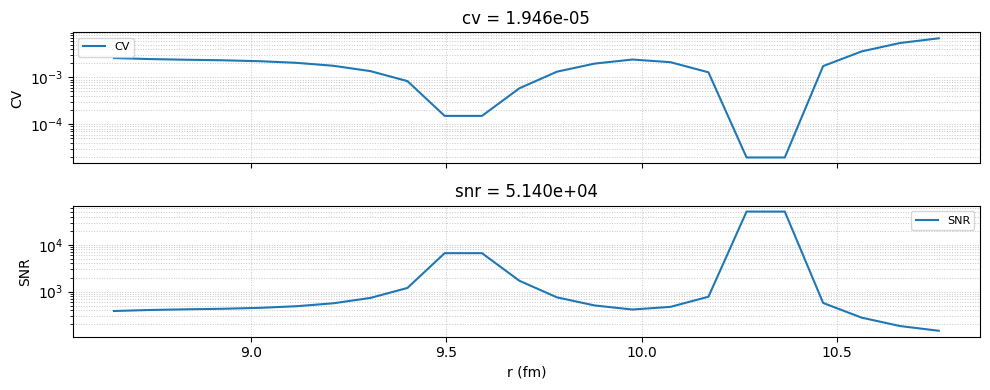}
\caption{\it The plot of the SNR (bottom) and CV (top) of the outgoing Jost function for the ground state using a window width of 3. The point of interest is the segment $r\in[10,10.5]$ which produced the highest (lowest) SNR (CV).}
\label{fig:snrcvforjostGroundState}
\end{figure*}  

\begin{figure*}[h!]
\centering
\includegraphics[width=0.85\textwidth]{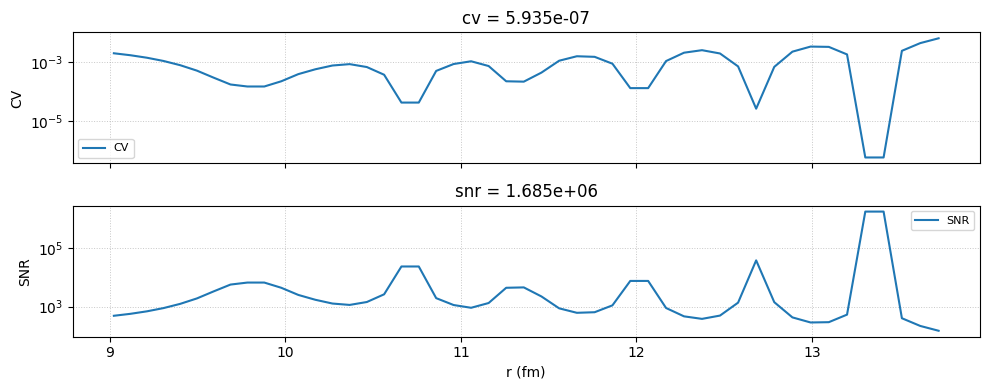}
\caption{\it The plot of the SNR (bottom) and CV (top) of the outgoing Jost function for the first excited state using a window width of 3. The point of interest is the segment $r\in[13,14]$ which produced the highest(lowest) SNR (CV).}
\label{fig:snrcvforjostfirstState}
\end{figure*}

\begin{figure*}[h!]
\centering
\includegraphics[width=0.85\textwidth]{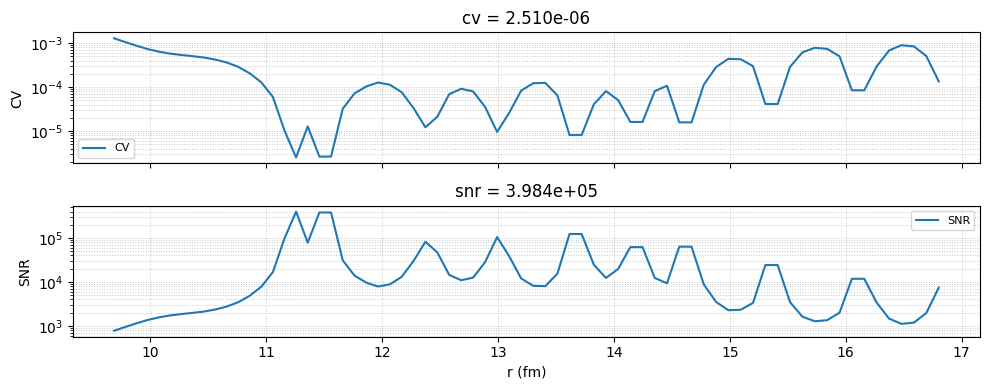}
\caption{\it The plot of the SNR (bottom) and CV (top) of the outgoing Jost function for the second excited state using a window width of 3. The point of interest is the segment $r\in[11,12]$ which produced the highest(lowest) SNR (CV).}
\label{fig:snrcvforjostsecondState}
\end{figure*}

\begin{figure}[!htb]
\minipage{0.33\textwidth}
  \includegraphics[width=\linewidth]{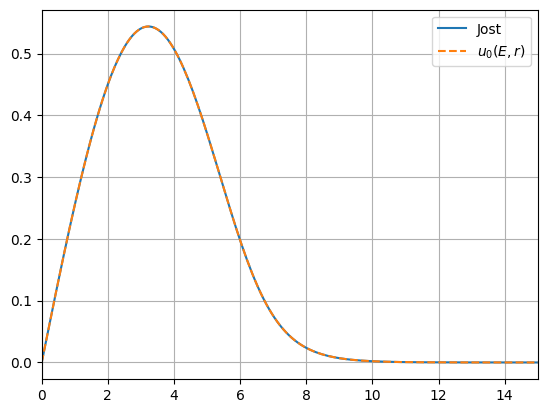}
\endminipage\hfill
\minipage{0.33\textwidth}
  \includegraphics[width=\linewidth]{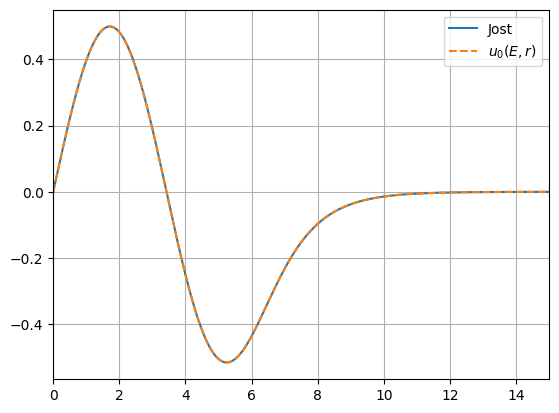}
\endminipage\hfill
\minipage{0.33\textwidth}%
  \includegraphics[width=\linewidth]{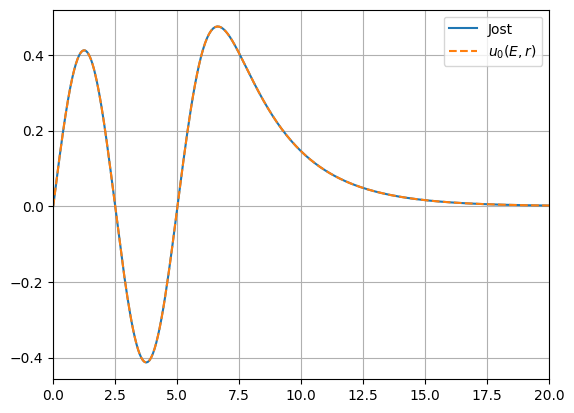}
\endminipage
 \caption{ Comparison of the ground (left), first excited (middle) and second excited (right) s-wave bound state wave functions obtained using Schrodinger equation Eq.~(\ref{singleChannelSE_SWave}) (orange) and asymptotic equation Eq.~(\ref{eq:wavefunction_asymptotic_with_jost}) (blue)}
\label{fig:comparisons_of_wavefunctions_with_js}
\end{figure}

In Fig.~\ref{fig:BiaseAndVarianceNormalResidue}, the bias squared is approximately three times larger than the variance term, indicating a high-bias, low-variance regime. This behaviour shows that the residual loss term systematically converges, on average, to an incorrect bound-state spectrum, as indicated by the violation of the expected spectral ordering, the appearance of a positive-energy state where a bound-state energy is expected to be negative, and the disagreement with the reference spectrum. Consequently, the eigenvalues obtained using this loss function are precise but inaccurate, meaning that the model consistently produces similar yet incorrect results.

In contrast, Fig.~\ref{fig:BiaseAndVarianceNormalizedResidue} shows that when the unnormalized residual loss is replaced by the normalized residual term, both the bias squared and the variance are reduced to values close to zero. This indicates that the predicted eigenvalues are both precise and accurate. Therefore, on average, the normalized residual loss recovers the correct physical solution and exhibits robustness with respect to the choice of initialization seed.

It is further observed that the normalized residual loss improves the learning of the ground-state eigenvalue, as both the bias squared and the variance attain their smallest values for this state. Moreover, the variance of the ground state remains approximately three times larger than the bias squared. For the excited states, the bias squared term dominates the error, consistent with the behaviour observed in Fig.~\ref{fig:BiaseAndVarianceNormalResidue}.

The fact that four independent random initializations converge to the same correct eigenvalues is consistent with the presence of a dominant physically meaningful basin of attraction associated with the true Schr\"odinger bound-state spectrum. In this case, the governing constraints (namely the differential equation, boundary conditions, and orthogonality conditions) are effectively balanced, and no evidence of spurious minima within the explored seed space and optimization settings. As a result, the model converges to the same physical solution across different initializations.

For the present setup, the normalized residual loss therefore provides the most reliable eigenvalues and is subsequently adopted for the computation of the Jost functions, as well as the residue of the $S$-matrix, the asymptotic normalization coefficient (ANC), and the nuclear vertex constant (NVC). This ensures that all subsequent scattering observables are constructed from a bound-state spectrum that is both numerically stable and physically consistent.

The convergence history of the predicted bound-state energies is shown in Fig.~\ref{fig:eigenvalueHistory}. The eigenvalues obtained from the Rayleigh–Ritz quotient converge within fewer than $10^3$ iterations out of the total $10^5$ training iterations for all three $s$-wave bound states, indicating rapid convergence of the method. Beyond approximately 1000 iterations, the predicted energies remain unchanged, and this regime is therefore not shown. The convergence histories corresponding to the other initialization seeds are omitted, as they exhibit similar behaviour.

Fig.~\ref{fig:WaveFunsWoods} shows the Woods–Saxon potential together with the three $s$-wave bound-state wave functions. The wave functions display the expected nodal structure, with the number of nodes increasing with excitation level, consistent with standard quantum-mechanical behaviour.

The stable bound-state solutions obtained with the normalized residual loss are used next to compute the Jost functions and the associated nuclear quantities. Figs.~\ref{fig:jostfuntionGroundState}, \ref{fig:finfoutFirstExcitedState}, and~\ref{fig:finfoutSecondExcitedState} show the computed outgoing Jost functions for the ground, first excited, and second excited states, respectively. The ground-state outgoing Jost function exhibits stronger numerical fluctuations than those of the excited states, indicating a larger level of numerical noise. By contrast, the second excited state shows the smallest variations, suggesting that its outgoing Jost function provides the most stable input for extracting the residue of the $S$-matrix, the ANC, and the NVC.

Since these derived quantities depend directly on the behaviour of the outgoing Jost function, a stable segment of its imaginary part must be selected before the residue, ANC, and NVC are evaluated. To identify this segment, sliding windows with widths in the range $[2,10]$ are applied, and the stability of each window is assessed using the signal-to-noise ratio (SNR) and the coefficient of variation (CV). Among the tested window sizes, a width of three gives the highest SNR and the lowest CV for all three states. This window is therefore used to extract the nuclear quantities. The corresponding SNR and CV results are shown in Figs.~\ref{fig:snrcvforjostGroundState}, \ref{fig:snrcvforjostfirstState}, and~\ref{fig:snrcvforjostsecondState}.

The residue of the $S$-matrix, ANC, and NVC, averaged over the four random seeds, are reported in Table~\ref{tab:meanEigenvaluesandstd}. The observed standard deviations follow the same trend as the fluctuations in the outgoing Jost functions: the ground state has the largest standard deviation, while the second excited state has the smallest. This indicates that the uncertainty in the extracted quantities is mainly associated with the numerical noise and numerical sensitivity (including discretization, sampling, and floating-point effects) introduced during the computation of the Jost function. Nevertheless, the theoretical values lie within one standard deviation of the computed mean values, demonstrating good agreement over the four independent random initializations.

It is important to emphasize that the Jost functions defined in Eq.~(\ref{def_of_jost}) are formally valid in the asymptotic region where the potential vanishes. In this region, the Jost functions are expected to be constant. Inside the potential region, however, this asymptotic definition is no longer guaranteed to apply in the same way. This behaviour is visible in Figs.~\ref{fig:jostfuntionGroundState}, \ref{fig:finfoutFirstExcitedState}, and~\ref{fig:finfoutSecondExcitedState}, where the Wronskian expressions used to determine the Jost functions show a clear dependence on the radial coordinate inside the potential region. This radial dependence indicates that this region is outside the formal asymptotic domain in which the Jost functions are defined as constants.

To assess whether this position-dependent behaviour still produces physically meaningful wave functions, the bound-state wave functions are reconstructed using Eq.~\ref{eq:wavefunction_asymptotic_with_jost}. The comparison with the original PINN wave functions is shown in Fig.~\ref{fig:comparisons_of_wavefunctions_with_js}. Although Eq.~(\ref{eq:wavefunction_asymptotic_with_jost}) is expected to be applicable primarily in the asymptotic region $\{r:V(r)\approx 0\}$, the reconstructed wave functions are visually indistinguishable from the original PINN wave functions over the plotted computational domain. This result shows that the PINN-based representation is able to reproduce the wave-function behaviour not only in the asymptotic region, but also in the interior region where the potential is non-zero. This observation suggests that the PINN-based Jost-function construction retains sufficient information to reconstruct the bound-state wave functions across the full domain. However, this result should be interpreted as numerical evidence rather than a formal proof of analytic continuation or of the analytic properties of the Jost function inside the potential region. Within this limitation, the agreement observed in Fig.~\ref{fig:comparisons_of_wavefunctions_with_js} indicates that the PINN-defined Jost functions provide a consistent route for extracting the residue of the $S$-matrix, ANC, and NVC from the computed bound-state solutions.

Overall, the predicted bound-state energies are in very good agreement with the theoretical values reported in Refs.~\cite{blokhintsev2007vertex,Mmusi}. The extracted residue of the $S$-matrix, NVC, and ANC are also in reasonable agreement with the reference values, although their standard deviations are larger than those obtained for the energies. This larger spread is consistent with the additional numerical sensitivity introduced through the Jost-function calculation. The RMS radii also agree closely with the available theoretical results, further supporting the consistency of the proposed PINN-based approach.

\section{Conclusion and future work}\label{sec:concl}
In this work, the applicability of the physics-informed neural network (PINN) framework to bound-state quantum problems and to the extraction of quantities related to the analytic structure of the scattering matrix was investigated. The method was applied to the $s$-wave bound-state Schr\"odinger equation for a Woods-Saxon potential without spin-orbit coupling. In the proposed formulation, the radial wave function was represented by an ANN, while the bound-state energies were obtained using the Rayleigh-Ritz variational quotient. From the resulting eigenpairs, the Jost functions were constructed using the Wronskian of the incoming/outgoing spherical solutions and the bound-state wave function, thereby providing access to quantities associated with the pole structure of the partial-wave $S$-matrix.

The main contribution of this work is a seed-robust residual-loss selection framework for PINN-based quantum bound-state calculations. Instead of selecting a residual loss from a single training run, two residual formulations were compared across four independent random seeds. Their performance was assessed using eigenvalue accuracy, coefficient of variation, signal-to-noise ratio, bias-variance decomposition, and Hermitian spectral-ordering consistency, where the latter refers to whether the predicted bound states preserve the expected ordered spectrum of the Hermitian Schr\"odinger operator. The normalized residual loss was found to provide the most stable and physically consistent results for the present setup. In contrast, the unnormalized residual loss could converge to an incorrectly ordered spectrum containing a spurious positive-energy solution. With the normalized residual loss, the predicted bound-state energies were in very good agreement with the theoretical values for the four random seeds considered. The RMS radii also agreed closely with the available theoretical results and exhibited very small standard deviations, indicating that the normalized residual loss stabilizes both the eigenvalues and the corresponding wave-function properties.

The stable bound-state solutions were then used to compute the Jost functions, the residue of the $S$-matrix, the asymptotic normalization coefficient, and the nuclear vertex constant. The extracted residue, ANC, and NVC showed reasonable agreement with the corresponding theoretical values, although their standard deviations were larger than those observed for the eigenenergies and rms radii. This larger spread was associated with the numerical sensitivity of the Jost-function calculation. In addition, the wave functions reconstructed from the Jost-function-based expression showed almost exact agreement with the original PINN wave functions over the full computational domain, suggesting that the PINN-based representation retains sufficient structural information beyond the asymptotic region. This observation should, however, be interpreted as numerical evidence rather than as a formal proof of analytic continuation.

Future work will focus on developing a more rigorous theoretical understanding of why the normalized residual loss improves stability in the PINN optimisation landscape, reducing the numerical uncertainty in the extracted residue, ANC, and NVC, and extending the framework to more general few-body and nuclear scattering applications.

\section*{Acknowledgments}
JTT gratefully acknowledges the Junior Research Fellowship(JRF) from the Department of Higher Education in South Africa. ASC is partially supported by the National Research Foundation of South Africa.

\bibliographystyle{IEEEtran}
\bibliography{biblio}

\end{document}